\documentclass{aa} 

\usepackage{graphicx}
\usepackage{txfonts}
\begin{document} 

   \title{
Scaling of the photon index vs mass accretion rate correlation  and estimate of black hole mass in M101 ULX-1
}


   \author{Lev Titarchuk
          \inst{1}
\and
          Elena Seifina\inst{2}
          }

   \institute{Dipartimento di Fisica, Universit\`a di Ferrara, Via Saragat 1, I-44122 Ferrara, Italy, 
\email{titarchuk@fe.infn.it}; \\ 
National Research Nuclear University MEPhI (Moscow Engineering Physics Institute), Moscow, Russia; 
Goddard Space Flight Center, NASA,  code 663, Greenbelt  
MD 20770, USA; \\ 
\email{lev@milkyway.gsfc.nasa.gov}, USA
%
         \and
Moscow State University/Sternberg Astronomical Institute, Universitetsky 
Prospect 13, Moscow, 119992, Russia; \\
\email{seif@sai.msu.ru
}             
             }

   \date{Received 
        ;	accepted  on  October 23, 2015
}


%


 
  \abstract
{We report the results of $Swift$ and $Chandra$ observations of an ultra-luminous X-ray source, ULX-1 in M101. We show   strong observational evidence that M101 ULX-1 undergoes  spectral transitions from the low/hard  state to the high/soft state during these observations. The spectra of M101 ULX-1 are well fitted by the so-{\it called} bulk motion Comptonization (BMC) model  for all spectral states. 
We have established the photon index ($\Gamma$) saturation level,
$\Gamma_{sat}$=2.8$\pm$0.1, in the $\Gamma$ vs.  mass accretion rate ($\dot M$) correlation. 
This $\Gamma-\dot M$ correlation allows us to evaluate  black hole (BH) mass in M101 ULX-1 to be  
$M_{BH}\sim (3.2 - 4.3)\times 10^4 M_{\odot}$ assuming the spread in distance  to M101 (from $6.4\pm 0.5$ 
Mpc to $7.4\pm 0.6$ Mpc). For this BH mass estimate we use the scaling method taking Galactic BHs  XTE~J1550-564, 
H~1743-322 and 4U~1630-472 as reference sources. The $\Gamma$ vs. $\dot M$ correlation revealed in M101~ULX-1 
is similar to that in a number of Galactic BHs and exhibits clearly the correlation along with 
the strong $\Gamma$ saturation at  $\approx2.8$. This is   {\it robust} 
observational evidence for the presence of a BH in M101 ULX-1. We also find that  the  seed (disk) photon 
temperatures are quite low, of order of 40$-$100 eV which is consistent with  high BH mass in M101~ULX-1. 
{Thus, we suggest that the central object in M101 ULX-1 has  intermediate BH mass of order 10$^{4}$ solar masses.} 
}


   \maketitle
%

\section{Introduction}

The Pinwheel Galaxy (also known as Messier 101, M101)
is a face-on spiral galaxy located  6 Mpc away 
in the constellation Ursa Major \citep{Shappee11}. 
At this distance an Earth observer can see only very bright sources 
whose X-ray luminosity is greater than $10^{38}$ erg s$^{-1}$ using current X-ray detectors. 
This galaxy has 
{\it
ten} 
ultra-luminous X-ray  (ULXs) sources 
[\cite{Pence01}]. 
M101 ULX-1 and ULX N5457-X9 
{are among them, 
}
which are well seen in X-rays. M101 ULX-1 was discovered 
with  {\it ROSAT} and identified as a ULX-1 
by \cite{Pence01}. 
The  bolometric  luminosity is in the range of $10^{40}-10^{41}$ ergs s$^{-1}$. Later, $Chandra$ observations 
(see Mukai et al. 2003; Di Stefano \& Kong 2003; Kong et al. 2004) found  a very $soft$ X-ray spectrum of  
this source with a blackbody temperature of about 100 eV. The source showed the low/hard 
and high/soft states in a quasi-recurrent manner  during 160$-$190  day period as found  by $Chandra$ 
and XMM-$Newton$ observations (Mukai et al. 2005). 

Two scenarios for interpretation of  ULX phenomena have been proposed. 
 First, these sources 
could be stellar-mass black holes [significantly less than 100 solar masses ($M_{\odot}$)] radiating at Eddington or super-Eddington rates 
[\cite{tl97}, \cite{Mukai05}].
Alternatively, they could be intermediate-mass black holes (IMBH, more than 100 $M_{\odot}$) where the luminosity is essentially 
sub-Eddington.
The exact  origin of such objects still remains uncertain. 

Given the faintness of the optical 
counterpart (typically V $>$ 22 mag; see for example Liu et al. 2004 and Roberts et al. 2008), radial velocity 
studies of ULX-1 have mostly concentrated on strong emission lines in the optical spectrum. However, these attempts to 
provide  a dynamical mass estimate of M101 ULX-1 {\it fail} because 
the emission lines are presumably  associated  with the accretion 
disk or a wind,  instead of the donor star itself (cf. Liu et al. 2012; Roberts et al. 2011). 
\cite{Mukai05} and \cite{Liu13}  estimated  BH mass in the range of 20$-$40 $M_{\odot}$ using the maximum of the bolometric luminosity for X-ray observations  
by $Chandra$ and XMM-$Newton$ during the high state. On the other hand, the estimates using the  dynamical method based on the  optical emission band provided quite a broad BH mass range. For example,  Liu et al. (2013) used optical 
HST observations of M101 ULX-1 to estimate  dynamical BH mass
in a wide range of $M_{BH}\sim 5-1000~M_{\odot}$.

The aformentioned BH mass evaluation, however contradicts with a relatively low  seed (disk) photon temperature of the blackbody  part of the spectrum which is in the range of 40-70 eV. 
For example, Shakura \& Sunyaev, (1973) (see also Novikov \& Thorne, 1973) give an effective temperature of the accretion material of $kT_{eff}\propto M_{BH}^{-1/4}$.
It is  desirable to have  an independent BH identification for the compact object  located  in the center  of M101 ULX-1 as an alternative to the dynamical method.

A new method of  BH mass determination was developed by Shaposhnikov \& Titarchuk (2009), hereafter ST09, using a correlation scaling between X-ray spectral and timing (or mass accretion rate) properties observed from many Galactic BH binaries during 
the spectral  state transitions. It is possible to evaluate a BH mass applying this method 
 when conventional dynamical methods cannot be used. 

{
Mukai et al. (2005),  Kong et al. (2004),  Kong \& Di Stefano (2005), have analyzed the $Chandra$ { and} 
XMM-$Newton$ spectra. They 
fitted the low/hard state ($L_X\sim 2\times 10^{37}$ erg s$^{-1}$) spectra with a power-law model, but they 
used a different model to fit the high/soft state spectra during the outbursts. Particularly,  Kong et al. fitted the outburst spectra with  the absorbed blackbody model of $kT_{BB}$=40$-$150 eV and $N_H = (1-4)\times 10^{21}$ cm$^{-2}$, 
and obtained outburst bolometric luminosities up to $3\times 10^{40}$ erg s$^{-1}$. In contrast, Mukai et al. fitted the spectra 
with a model consisting of a blackbody plus a diskline component centered at 0.5 keV with $N_H$ fixed at $4\times 10^{20}$ cm$^{-2}$, or with the absorbed 
blackbody with $N_H$ ranging from 0.4 to 3.7$\times 10^{21}$ cm$^{-2}$. Note Liu (2009), based on HST observations,  indicates  a smaller 
absorption in the  range $(1-6)\times 10^{20}$ cm$^{-2}$.
Thus the absorbing column $N_{H}$ for M101  is in  a wide range 
depending on different X-ray and optical observations and also assuming various emission  models of the source.

  \begin{figure*}
    \includegraphics[width=15cm]{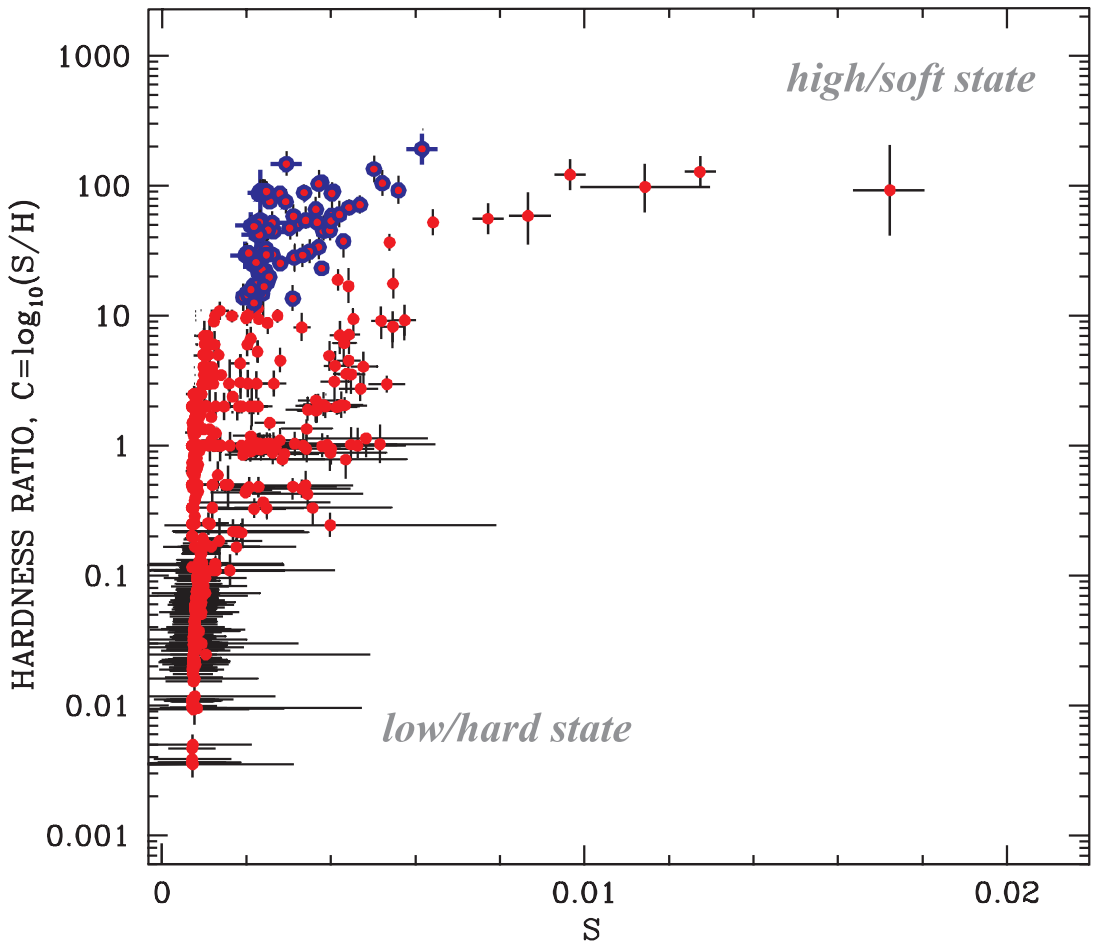}
   \caption{
Color-intensity evolutionary diagram for M101 ULX-1 using $Swift$ observations (2006 --2012), where S and H 
are the source counts in the two bands: the $soft$ [0.3 -- 1.5 keV]  and $hard$ [1.5 -- 10 keV] passbands. 
Spectral softness (hardness ratio) C and  source soft  brightness $S$ increase 
to the right and  finally the former one  saturates  at high values of S.   
Larger values of $C$ indicate a softer spectrum, and vice versa. 
{
$Blue$ points are related to  a decay part of the outburst (see  the light curve in  Fig.~\ref{lc}.
}
}
   \label{HID}
 \end{figure*}

As for the distance estimate for M101  ULX-1, Kelson et al. (1996) 
provide  a value  of 7.4 Mpc while Freedman et al. (2001) argue that the distance is less and it is  
about 6.8 Mpc. 
}
Recently, 
Shappe \& Stanek (2011) obtained 
a Cepheid distance to M101 using archival HST/ACS 
time series photometry of the $inner$ fields of 
the galaxy based on  a larger Cepheid sample. 
They improved the distance  determination for M101  and obtained a distance value of 
$d_{m101}=6.4\pm$0.5 Mpc. 

In this Paper we present an analysis of  available 
{\it Swift} and  
$Chandra$  observations of M101 ULX-1. 
In \S 2 we present the list of observations used in the data analysis while 
in \S 3 we provide  details of the X-ray spectral analysis.  We discuss the evolution of 
the X-ray spectral properties during the high-low state  transition 
and present the results of the scaling analysis to estimate BH mass of M101 ULX-1 in \S 4. 
We make our final  conclusions in \S 5. 

%
%

  \begin{figure*}
 \centering
    \includegraphics[width=17cm,clip]{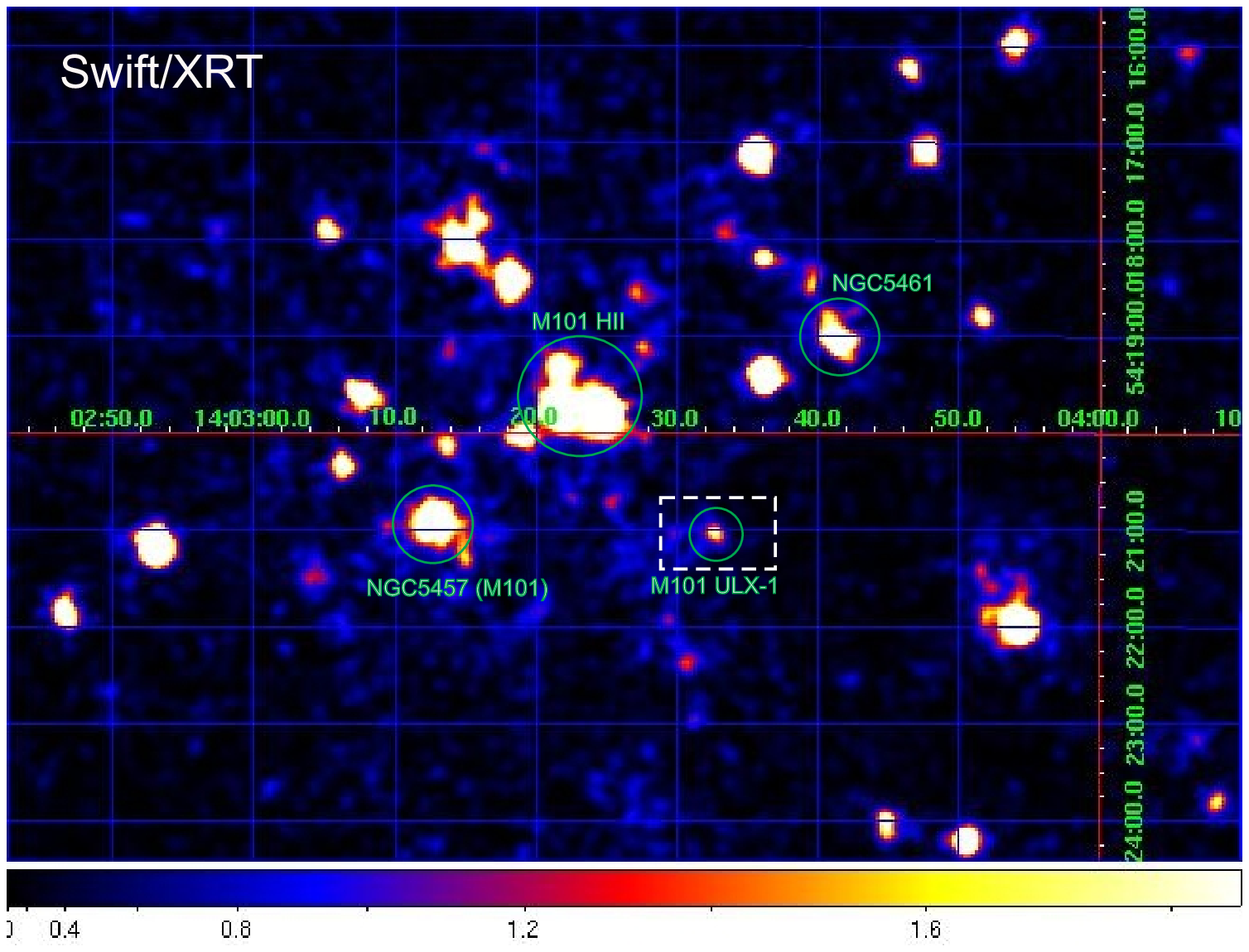}
      \caption{$Swift$/XRT (0.3$-$10 keV) image of M101 field of view, where
green circles are the locations of M101 ULX-1, M101 galactic nucleus (NGC 5457), 
NGC 5461 and M101 H II regions.
The image segment selected by dashed line box is also shown in Fig.~\ref{imageb} using $Chandra$ data.
}
      \label{imagea}
 \end{figure*}

%
%

  \begin{figure*}
 \centering
    \includegraphics[width=17cm]{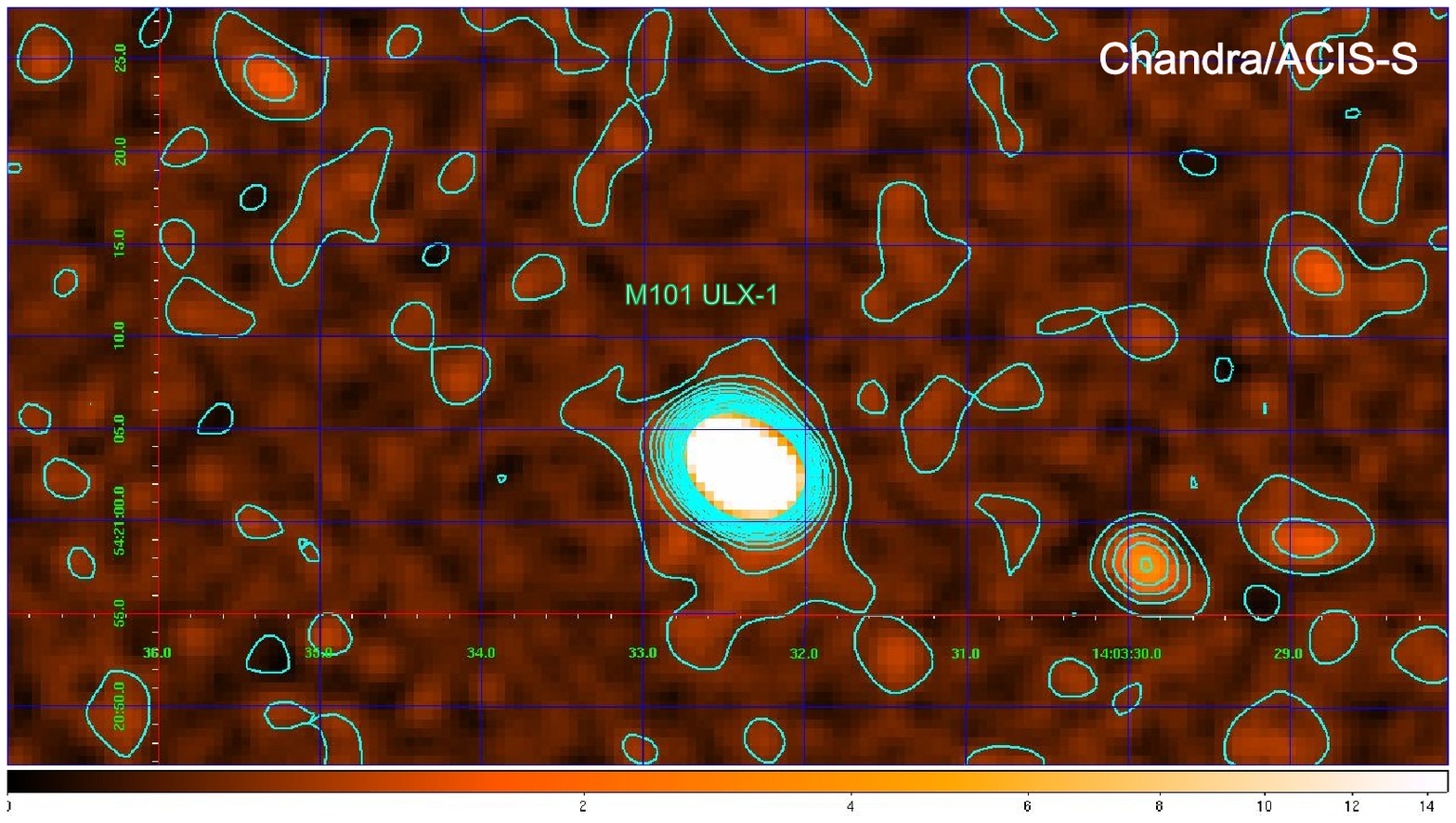}
      \caption{
Adaptively smoothed $Chandra$/ACIS-S (0.2-8 keV) image  of the M101 field, 
which localization is indicated by the selected  
{dashed line}
box in Fig.~\ref{imagea}.
 Contour levels 
demonstrate the minimal contamination by other point sources and diffuse emission 
within circle of 9{\tt "} around M101 ULX-1. 
}
\label{imageb}
\end{figure*}

\section{Observations and data reduction \label{data}}

{
As the first step  we analyzed  the {\it Swift} data set for M101 ULX-1, which covered the longest observational interval (2006 -- 2013). 
In this way, we studied the source behavior in  X-rays (Sect.~\ref{swift data}). \
Then we proceeded with a detailed spectral analysis using the {\it Chandra} (2000, 2004 -- 2005) 
data 
(\S~\ref{chandra data}). A summary 
of the X-ray observations considered in this work is given in Tables 1 and 2.    


\begin{table*}
 \caption{The list of $Swift$ observations of M101 ULX-1 used in our analysis}              
 \label{tab:list_Swift}      
 \centering                                      
 \begin{tabular}{l l l l l c}          
 \hline\hline                        
  Obs. ID& Start time (UT)  && End time (UT) &MJD interval \\    
 \hline                                   
00035892001       & 2006 Aug. 29 11:38:56 && 2006 Aug. 29 21:24:57 & 53976.8 -- 53976.9 &\\
  00030896(001-009) & 2007 March 1          && 2007 Apr. 19          & 54160 -- 54209 &\\
  00032081(001-149) & 2011 Aug. 24          && 2012 May 10           & 55797 -- 56058 &\\
  00032094(001-018) & 2011 Sep. 7           && 2013 Sep. 11          & 55811 -- 56546 &\\
  00032101(001-013) & 2011 Sep. 23          && 2013 Sep. 20          & 55827 -- 56555 &\\
  00032481001       & 2012 June 9 10:17:15  && 2012 June 9 13:55:57  & 56087.4 -- 56087.5&\\
 \hline                                             
 \end{tabular}
 \end{table*}

\subsection{{\it Swift} data\label{swift data}}

The log of the {\it Swift}/XRT observations used in this Paper  is shown in Table 1. 
The {\it Swift} source count rates never exceed 0.02 count s$^{-1}$, therefore  only photon-counting mode (PC) events (selected in grades 0$-$12) 
were considered. In this way, the $Swift$-XRT/PC data (ObsIDs, indicated in the
first column of 
Table 1) 
were processed using the HEA-SOFT v6.14, the tool XRTPIPELINE v0.12.84 and the
calibration files (CALDB version 4.1).
The ancillary response files were created using XRTMKARF v0.6.0 and exposure maps generated by XRTEXPOMAP v0.2.7. 
We fitted the spectrum using the response file SWXPC\-0TO12S6$\_$20010101v012.RMF.
We also used the online XRT data product generator\footnote{http://www.swift.ac.uk/user\_objects/}  for independent check: 
light curves and spectra (including background and ancillary response files, 
see Evans et al. 2007, 2009). 
We have 
made the state identification in terms of the color ratio (see Sect.~3.2), 
using the Bayesian method developed  by Park et al. (2006).
Moreover,  we have applied the effective area option of 
the Park's code which  includes the count-rate correction factors in their calculations.  
Our results, adapting this technique,  
indicate to  two color$-$intensity regimes in M101 ULX-1: 
i. with low color ratio at lower count-rate observations and ii. high color ratio at higher count events (see Figure \ref{HID}).
Furthermore the color$-$intensity diagram shows a smooth track. Therefore, 
%
we have grouped the $Swift$ spectra into four bands according to count rates (see Sect.~3.1) 
and fitted the combined  spectra of each band using the {\tt XSPEC} package (version 12.8.14).
 

%
%


%
%

  \begin{figure*}
 \centering
    \includegraphics[width=17cm]{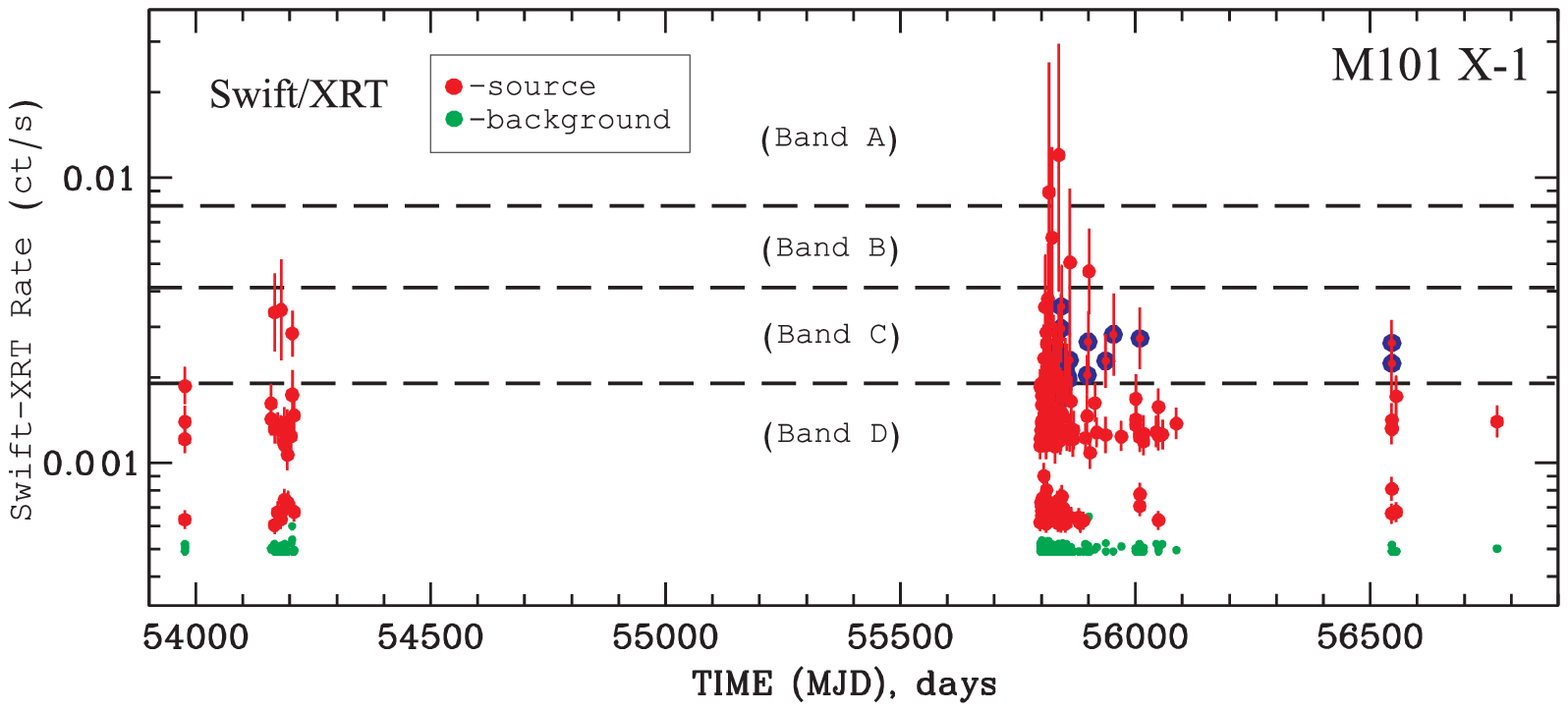}
      \caption{
$Swift$/XRT light curve of M101 ULX-1 in the 0.3$-$10 keV band during 2006 -- 2013.  
$Red$ points mark the source signal 
and $green$ points indicate the background level.
{
$Blue$ points mark decay part of the outburst for this light curve and correspond to $blue$ points of color-intensity 
evolutionary diagram  (see higher branch) shown in Fig.~\ref{HID}.
}
}
   \label{lc}
 \end{figure*}

\begin{table*}
 \caption{The list of $Chandra$ observations of M101 ULX-1 }
 \label{tab:list_Chandra}
 \centering 
 \begin{tabular}{l l l l l l l l l}
 \hline\hline                        
 Obs. ID & Start time (UT) & Rem. & Obs. ID & Start time (UT)& Rem. & Obs. ID & Start time (UT) & Rem. \\
 \hline                                   
934$^{1,2,3,4}$ & 2000-03-26 & HS & 5322$^{1,3}$ & 2004-05-03 & LS & 4734$^{1,3}$ & 2004-07-11 & HS \\
2065$^{1,2,3}$& 2000-10-29 & HS & 4733$^{1,3}$ & 2004-05-07 & LS & 4736$^{1}$   & 2004-11-01 & LS \\
4731$^{1,3}$  & 2004-01-19 & LS & 5323$^{1,3}$ & 2004-05-09 & LS & 6152$^{1}$   & 2004-11-07 & LS \\
5297$^{1,3}$  & 2004-01-24 & LS & 5337$^{1,3}$ & 2004-07-05 & HS & 6170$^{1,2}$ & 2004-12-22 & LS \\ 
5300$^{1,3}$  & 2004-03-07 & LS & 5338$^{1,3}$ & 2004-07-06 & HS & 6175$^{1,2}$ & 2004-12-24 & LS \\
5309$^{1,3}$  & 2004-03-14 & LS & 5339$^{1,3}$ & 2004-07-07 & HS & 6169$^{1,2}$ & 2004-12-30 & HS \\
4732$^{1,3}$  & 2004-03-19 & LS & 5340$^{1,3}$ & 2004-07-08 & HS & 4737$^{1,2}$ & 2005-01-01 & HS \\
 \hline                                             
 \end{tabular}
 \tablebib{
(1) Mukai et al. 2005; 
(2) Kong \& Di Stefano 2005; 
(3) Kong et al. 2004; 
(4) Pence et al. 2001.}
\tablefoot{ 
HS and LS are related to {\it high state/low state} of M101 ULX-1.}
 \end{table*}

\subsection{{\it Chandra} data \label{chandra data}}


M101 ULX-1 was also observed by {\it Chandra} in 2000, 2004$-$2005. 
The log of {\it Chandra} observations used in this Paper  is presented 
in Table 2. 
We extracted spectra from the ACIS-S detector
using the standard pipeline CIAO v4.5 package and calibration database CALDB 2.27.
All data were taken in very faint mode (VFAINT) except for  the data taken in 2000,  March 26  and October 29, which were 
used in faint mode (FAINT). 
We have also identified  intervals of high background level in order to exclude all 
high background events. 
 The {\it Chandra} spectra were produced and  
modelled over the 
0.3 -- 7.0 keV energy range. Note that the data during the {\it low state} 
(indicated by * in Table \ref{tab:list_Chandra}), 
are characterized by only  a few  photons 
($\sim$10 -- 30)  for each  observation. Therefore, we combined all  the {low state} data to perform statistically 
significant spectral fits. 
Thus, we present the results for
these {low state} data per observation 
using C-statistic. While the rest of the data are analyzed in terms of $\chi^2$-statistics.

\section{Results \label{results}}

\subsection{Images \label{image_lc}}

In order to avoid a possible contamination from nearby sources we 
made a visual inspection 
of the obtained image  (smoothed by a Gaussian with an FWHM of 3{\tt "}2). 
{ Swift}/XRT (0.3 -- 10 keV) image of M101 field of view 
is presented in 
Figure~\ref{imagea}, where  {\it green} 
circles are the locations of M101 ULX-1, NGC~5457 (M101), 
NGC~5461 and M101 H~II regions. 
 
For  deeper image analysis 
we used the {\it Chandra} images with better data quality, 
provided by ACIS-S onboard {\it Chandra}. 
We point out the {\it Chandra} region as shown 
{by dashed line 
}
 box in the {\it Swift} image in 
Figure~\ref{imagea}.  
The {\it Chandra}/ACIS-S (0.2-8 keV) image obtained during observations of M101 ULX-1 on March 26, 2000 
(with exposure time of 99.5 ks, ObsID=934) is diplayed in Figure \ref{imageb}. 
 Contour levels 
should demonstrate the minimal contamination by other point sources and diffuse emission 
within circle of 9 arsec around M101 ULX-1.
For each observation, we extracted the source spectrum from a 9{\tt "} radius
circular region centered on the source position of M101 ULX-1 [$\alpha=14^{h}03^{m}32^s.37$, 
$\delta=54^{\circ} 21{\tt '} 02{\tt ''}.7$, J2000.0, see details in \cite{Kuntz05}], while an annulus region 
centered on the source
 with 10 and 18{\tt "} radii 
 was used to estimate  the background contribution.

 In addition, we extracted emission related to the other bright nuclear sources NGC~5457, NGC~5461 from  circular regions with radius of 
15{\tt "} and retrace their time behavior. 
As a result  we established that only M101 ULX-1 demonstrated  significant variability during the analyzed  observations. 

\subsection{Color-intensity diagrams and light curves  \label{HID_lc}}

Before detailed detailed spectral fitting  we investigated a so called color ratio to quantify and characterize the source spectrum. In particular, for our $Swift$ data we consider $R$ as a  ratio of the counts $S$ 
and $H$ in the soft (0.3 -- 1.5 keV) and hard (1.5 -- 10 keV) bands, respectively.
 However, at low counts, the posterior distribution of the counts ratio, $R$, tends to be skewed
because of the Poissonian nature of data. Therefore we used the color, $C=log_{10}(S/H)$, which  a log 
transformation of $R$, which provides the skewed distribution more symmetric (see e.g., Park et al. 2006).
The ratio $C$ is modified by taking into account background counts and instrumental effective areas. Figure~\ref{HID} demonstrates the color-intensity diagram and thus one can  see that different count-rate observations correspond to different color regimes. 
Larger values of $C$ indicate a softer spectrum, and vice versa. 
Note that  we have applied a Bayesian approach to compute the  ratio values $C$ and their errors using BEHRs software 
~\citep{Park06}\footnote{A Fortran and C-based program which calculates the 
ratios using the methods described by 
\cite{Park06} 
(see http://hea-www.harvard.edu/AstroStat/BEHR/)}.
Generally, this method is  applicable 
when the source is faint or the background is relatively large~\citep{Evans09,Burke13,Jin06}. In our case, the most $Swift$ observations are 
related to low count-rate regimes, which can confuse a reliable color estimates. 
However, Bayesian analysis provides a simple way to overcome  this 
problem.  As a  result, we found a clear LS-HS evolution of X-ray emission from M101 ULX-1. 
%
Furthermore, Figure~\ref{HID}   
demonstrates that the color $C$ monotonically increases with the soft flux $S$ and achieves  
a noticeable stability at high soft fluxes. 
{
Note that the color-color diagram of M101 ULX-1 clearly demonstrates two groups of datapoints,  related to the high/soft and low/hard states (see Fig. \ref{HID}). More specifically, in outbursts, M101 ULX-1 evolves from the 
$hard$ state to the $soft$ state during the rise phase and then returned to the $hard$ state during the decay phase. 
This evolution is similar to most outbursts of Galactic 
X-ray binary transients (e.g.
Homan et al. 2001; Shaposhnikov \& Titarchuk, 2006; Belloni et al. 2006; ST09; TS09; 
Shrader et al. 2010; Mu$\tilde n$oz-Darias et al. 2014).
}

The source M101~ULX-1 is 
in the {\it low state} (characterized by a low count rate) during most of the time  except for  
relatively short outbursts (with a high count rate, see   Fig.~\ref{lc} for details).  
Because of a low count rate we combined all of the {\it low state} data for {\it Chandra} 
and {\it Swift} data.   

In Figure~\ref{lc} we present {\it Swift}/XRT light curve of M101 ULX-1 during 2006 -- 2013 for the 0.3 -- 10 keV band. 
{\it Red} points mark the source signal 
and $green$ points indicate the 
background level. 
We have detected an outburst of M101 ULX-1 at MJD=55800 -- 56100, while for the rest of the
{\it Swift} observations this source remained 
in the {\it low state}. Individual {\it Swift}/XRT observations of M101 ULX-1 in PC ({\it Photon counting}) mode do not have 
enough counts to allow statistically meaningful spectral fits. To overcome 
this problem, we have examined the {\it Swift}/XRT lightcurve and grouped the observations into 
four bands: very high ("A"), high ("B"), medium ("C") and low ("D") count rates (see Fig.~\ref{lc}). 
{
We have also  split Band C into two subbands. Blue points shown in Figure~\ref{HID}  are associated with 
softer/higher track (see also related points in  
the  lightcurve, Fig.~\ref{lc}). 
In fact, this softer track ($blue$ points of Figure~\ref{HID}) corresponds to the outburst 
decay part (
see Fig.~\ref{lc}). 
While Band-C$_h$ (red points) are 
related to  the lower track of  the color-intensity diagram.
}
Finally, we  have combined 
the spectra  in each  related band and fitted them for all these observations
using $\chi^2$ statistics.
In addition, some of the brightest source spectra of A- and B-sets 
were regrouped with the task grppha and then
analysed in the 0.3 -- 7 keV  range 
using 
the Cash 
statistics.



\subsection{Spectral Analysis \label{spectral analysis}}


{
We examine  different spectral models in application to all available data  for M101 ULX-1
in order to describe the source evolution between the $low$ and $hard$ states.
Specifically, we use the combined $Swift$ 
spectra 
from  different spectral states to test a number of spectral models: 
$powerlaw$, $Bbody$, $bmc$ and their possible combinations modified 
by an absorption  model. 
{
}
We fitted all spectra using a tied neutral column, which provides the best-fit  column $N_H$ of $3\times 10^{21}$ cm$^{-2}$.

%
%

  \begin{figure*}
 \centering
    \includegraphics[width=17cm]{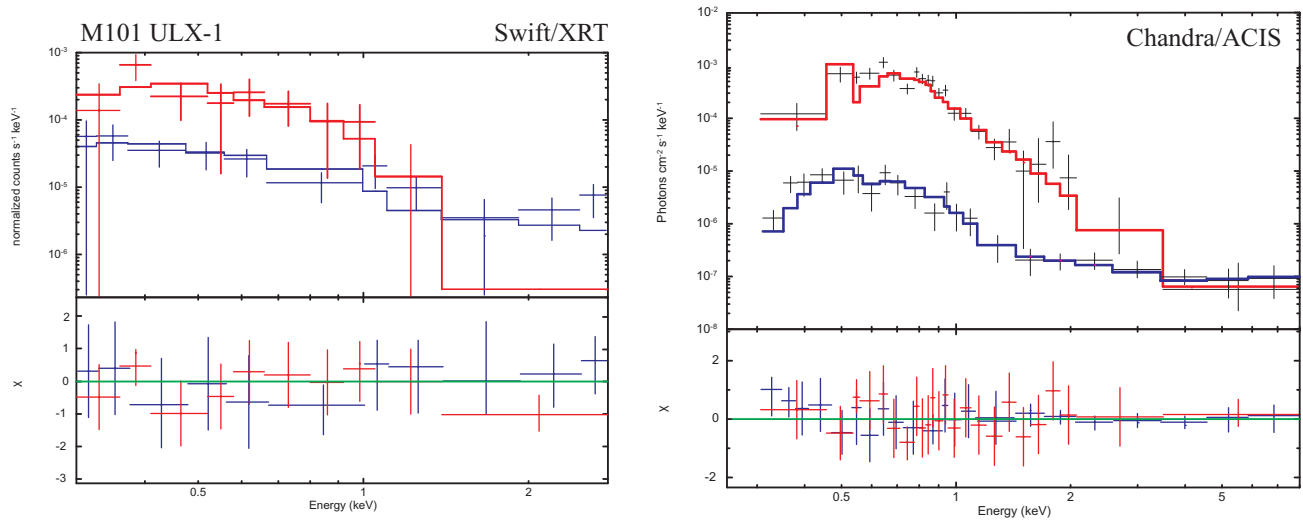}
      \caption{
($Left$:) $Swift$/XRT spectra from band A ($red$) and band C ($blue$) fitted using the $bmc$ model. 
($Right$:) 
two representative $Chandra$ spectra for different states of M101 ULX-1. Data  taken for  2004 July 5 ($red$), 
correspond to the  high state spectrum and for 2004 January -- May and November ($blue$), correspond to   the  
low state spectrum. 
Spectra are fitted by a $phabs*bmc$ model  with $kT_{bb}=70$ eV (red solid line), for  the high state and  with 
$kT_{bb}=45$ eV (blue solid line), for the  low state. See also the best-fit parameters 
listed in Tables~\ref{tab:par_swift} and \ref{tab:fit_table_Chandra} for $Swift$ and $Chandra$ data, respectively.
}
\label{two_state_spectra}
\end{figure*}


\begin{table*}
 \caption{Best-fit parameters  of the combined $Swift$ spectra
 of M101 ULX-1 in the 0.3$-$7~keV energy range using the following four 
models$^\dagger$: $phabs*power$, $phabs*bbody$, $phabs*(bbody+power)$ and $phabs*bmc$ 
}
    \label{tab:par_swift}
 \centering 
 \begin{tabular}{lllllll}
 \hline\hline
Model & Parameter & Band-A & Band-B & Band-C$_h$ & Band-C$_s$ & Band-D \\
 \hline                                             
Power-law   & $\Gamma_{pow}$    & 6.2$\pm$0.2 & 3.6$\pm$0.3 & 1.9$\pm$0.2  &  2.0$\pm$0.2  & 1.4$\pm$0.2 \\
            & N$_{pow}^{\dagger\dagger}$ & 2.8$\pm$0.03 & 1.4$\pm$0.02 & 0.67$\pm$0.05  & 0.68$\pm$0.04 & 0.04$\pm$0.01 \\
           & $\chi^2$ {\footnotesize (d.o.f.)} & 2.3 (18)     & 2.15 (18)     & 2.03 (18) & 2.02 (18)   & 1.15 (18) \\
      \hline
Bbody      & T$_{BB}$           & 65$\pm$2   & 70$\pm$3 & 85$\pm$3     & 84$\pm$5     & 94$\pm$4          \\
           & N$_{BB}^{\dagger\dagger}$ & 5.2$\pm$0.5 & 4.5$\pm$0.3    & 2.7$\pm$0.6 & 2.8$\pm$0.5 & 1.5$\pm$0.4 \\
           & $\chi^2$ {\footnotesize (d.o.f.)} & 1.14 (18)     & 1.28 (18)     & 1.94 (18)     & 1.93 (18)     & 3.03 (18) \\
      \hline
Bbody      & T$_{BB}$          & 70$\pm$3   & 86$\pm$4   & 90$\pm$5    & 89$\pm$3  & 70$\pm$4         \\
           & N$_{BB}^{\dagger\dagger}$ & 4.2$\pm$0.5 & 3.6$\pm$0.6    & 1.4$\pm$0.6   & 1.6$\pm$0.5 & 1.3$\pm$0.4 \\
Power-law  & $\Gamma_{pow}$    & 2.2$\pm$0.1 & 2.1$\pm$0.4 & 1.4$\pm$0.1   & 1.5$\pm$0.1 & 3.4$\pm$0.3 \\
           & N$_{pow}^{\dagger\dagger}$ & 0.64$\pm$0.01& 0.57$\pm$0.03 & 0.36$\pm$0.09    & 0.38$\pm$0.07 & 1.3$\pm$0.4 \\
           & $\chi^2$ {\footnotesize (d.o.f.)} & 1.23 (16)     & 1.19 (16)  & 1.23 (16)  & 1.22 (16) & 1.27 (16) \\
      \hline
bmc        & $\Gamma_{bmc}$    & 2.5$\pm$0.3    & 2.1$\pm$0.2  & 1.6$\pm$0.1    & 1.7$\pm$0.1    & 1.4$\pm$0.1 \\
           & T$_{s}$          & 92$\pm$10      & 76$\pm$9     & 56$\pm$10      & 57$\pm$10      & 42$\pm$8    \\
           & logA$$            & -5.3$\pm$0.4   & -4.7$\pm$0.5 & -4.3$\pm$0.4   & -4.2$\pm$0.5   & -3.9$\pm$0.5 \\
           & N$_{bmc}^{\dagger\dagger}$ & 15.6$\pm$0.5& 8.1$\pm$0.3 & 4.4$\pm$0.2  & 4.5$\pm$0.4  & 2.9$\pm$0.2 \\
           & $\chi^2$ {\footnotesize (d.o.f.)} & 1.21 (16) & 0.97 (16)  & 1.15 (16)&  1.14 (16)& 1.03 (16) \\
 \hline                                             
 \end{tabular}
\tablefoot{ 
$^\dagger$     Errors are given at the 90\% confidence level. 
$^{\dagger\dagger}$ The normalization parameters of $Blackbody$ and $bmc$ components are in units of $L^{soft}_{35}/d^2_{10}$ erg s$^{-1}$ kpc$^{-2}$, 
where $L^{soft}_{35}$ is the soft photon luminosity in units of $10^{35}$ erg s$^{-1}$, $d_{10}$ is the distance to the 
source in units of 10 kpc, and $Power$-$law$ component is in units of 10$^{-4}$ 
keV$^{-1}$ cm$^{-2}$ s$^{-1}$ at 1 keV. 
$N_H$ is the column density for the neutral absorber, $3\times 10^{21}$ cm$^{-2}$ (see details in the text). 
$T_{BB}$ and $T_{s}$ are the temperatures of 
the $blackbody$ and seed photon components, respectively (in eV). $\Gamma_{pow}$ and $\Gamma_{bmc}$ are the indices of the {\it power law} 
and $bmc$, respectively. 
}
 \end{table*}

\subsubsection{Choice of the Spectral Model\label{model choice}}

As a first step, we proceed with a model of an absorbed power-law.  This model ($phabs*powerlaw$) 
fits  well   the low state data only [e.g., for D-spectra, $\chi^2_{red}$=1.15 (18 d.o.f.), see 
{the  left column of }
Table~3]. As one can see the power-law model is characterized by  very large photon indices (much greater than 3, particularly for A and B-event spectra, see notations of these events in Fig. \ref{lc}) 
and furthermore, this model  gives unacceptable fits 
(e.g., for all A,  B and C-spectra of $Swift$ data). 
On the other hand, for the $high$ state data, the thermal model ($Bbody$) provides better fits 
than the power-law model. However, the intermediate state spectra  (B-, C-spectra for $Swift$ data)
cannot be fitted by any single-component model. In particular, a simple power-law model produces a soft excess. 
Significant positive residuals at low energies less than 1 keV suggest the presence of additional emission 
components. For this reason, we also use a sum of blackbody and power-law component model ($N_H=3\times 10^{21}$ cm$^{-2}$, 
$kT_{bb}=70-90$ eV, and $\Gamma=1.4 - 2.2$; see 
Table 3).
The best fits of 
$Swift$ 
spectra has been 
obtained by implementation of  the so called {\it Bulk Motion Comptonization model} [{\tt BMC XSPEC} model, \cite{tl97}],  
for which the photon index ranges from $\Gamma\sim 1.4 - 2.8$  for all observations (see Tables 3, 4 and Fig. \ref{two_state_spectra}). Furthermore, we achieve the 
best-fit results using the same model for all spectral ($high$ and $low$) states. 

%
%


{ 
{
}
}
{
%
%
}
We should remind a reader that  the BMC model is characterized by the seed photon  temperature $T_s$, the energy index of the Comptonization spectrum 
$\alpha$ ($\alpha=\Gamma-1$), 
the illumination parameter $\log(A)$ related to the Comptonized (illumination) fraction $f=A/(1+A)$. This model convolves a seed (disk) blackbody  with an upscattering   Green's function. 
{ We also use a multiplicative $phabs$ component  to take into account  an absorption by neutral material. The $phabs$ model parameter is an equivalent hydrogen column $N_H$. 
{In Table 3 we demonstrate a good performance of the BMC model in application to the $Swift$ data ($0.97<\chi^2_{red}<1.21$).}

\begin{table*}
\caption{Best-fit parameters of the spectra using  $Chandra$ observations of M101~ULX-1 in the 0.3 -- 7~keV 
energy range$^{\dagger}$. Parameter errors correspond to 90\% confidence level.}\label{tab:fit_table_Chandra}
\centering
\begin{tabular}{lcccccccc} 
\hline\hline
ObsID & MJD, day & Exp, ks & Counts & $kT_s,$ keV  & $\Gamma_{bmc}=\alpha_{bmc}+1$     & $\log{A}$ & N$_{bmc}^{\dagger\dagger}$ &  $\chi^2_{red}$ (d.o.f.), MC$^{\dagger\dagger\dagger}$ \\
\hline
934   & 51629 & 94  &   8642 &   100$\pm$21 &  2.78$\pm$0.08 & -3.78(9) & 35.2(3) & 0.99 (28) \\
2065  & 51846 & 10  &   310  &   67$\pm$10  &  2.6$\pm$0.1   & -2.36(8) & 18.9(1) & 1.08 (10) \\
4731* & 53023 & 56  &   26   &   46$\pm$10  &  1.39$\pm$0.07 & -2.1(5)  & 2.2(1)  &  0.89     \\
5297* & 53028 & 15  &   14   &   42$\pm$9   &  1.38$\pm$0.04 & -2.0(6)  & 2.3(2)  &  0.78     \\
5300* & 53071 & 52  &   13   &   43$\pm$8   &  1.38$\pm$0.05 & -2.0(6)  & 2.2(1)  &  0.99     \\
5309* & 53078 & 71  &   18   &   44$\pm$9   &  1.37$\pm$0.06 & -2.0(5)  & 2.1(1)  &  0.98     \\
4732* & 53083 & 70  &   12   &   42$\pm$8   &  1.38$\pm$0.04 & -2.0(3)  & 2.1(1)  &  0.91     \\
5322* & 53128 & 65  &   17   &   45$\pm$10  &  1.39$\pm$0.08 & -2.0(5)  & 2.2(1)  &  0.93     \\
4733* & 53132 & 16  &   12   &   41$\pm$7   &  1.36$\pm$0.07 & -2.0(4)  & 2.1(1)  &  0.85     \\
5323* & 53134 & 43  &   10   &   40$\pm$10  &  1.35$\pm$0.09 & -2.5(2)  & 2.0(1)  &  0.82     \\
5337  & 53191 & 10  &   129  &   70$\pm$12  &  1.65$\pm$0.09 & -3.32(9) & 4.6(2)  &  0.97 (12) \\
5338  & 53192 & 28  &   162  &   98$\pm$25  &  1.89$\pm$0.07 & -2.93(8) & 6.3(1)  &  1.00 (30) \\
5339  & 53193 & 14  &   468  &   65$\pm$14  &  1.97$\pm$0.1  & -4.18(6) & 6.9(1)  &  1.08 (20) \\
5340  & 53194 & 54  &   680  &   51$\pm$3   &  2.72$\pm$0.09 & -2.4(3)  & 30.6(3) &  1.21 (23) \\
4734  & 53197 & 35  &   582  &   60$\pm$9   &  2.12$\pm$0.06 & -3.9(4)  & 8.7(1)  &  1.25 (14) \\
4736* & 53310 & 78  &   29   &   45$\pm$8   &  1.36$\pm$0.07 & -2.4(2)  & 2.0(1)  &  0.89      \\
6152* & 53316 & 44  &   21   &   43$\pm$9   &  1.36$\pm$0.08 & -2.4(3)  & 2.1(2)  &  0.96      \\
6170  & 53361 & 48  &   41   &   47$\pm$12  &  1.5$\pm$0.1   & -2.0(1)  & 3.1(5)  &  0.6  (5)  \\
6175  & 53363 & 41  &   54   &   45$\pm$10  &  1.9$\pm$0.3   & -3.7(1)  & 5.7(1)  &  0.78 (6)  \\
6169  & 53369 & 29  &   613  &   71$\pm$5   &  2.1$\pm$0.1   & -4.1(2)  & 8.1(1)  &  1.12 (20) \\
4737  & 53371 & 20  &   1483 &   95$\pm$7   &  2.75$\pm$0.06 & -3.9(1)  & 26.7(1) &  1.08 (54) \\
Comb.LS** & ... & 500 & 172 & 45$\pm$10    &  1.39$\pm$0.08 & -2.0(5)  & 2.2(1)  &  0.93 (10) \\
\hline
\end{tabular}
\tablefoot{ 
$^\dagger$ The spectral model is  $phabs*BMC$, where $N_H$ is 
$3\times 10^{21}$ cm$^{-2}$ as a best-fit neutral absorption obtained for both $Chandra$ and $Swift$ spectra for the $low$ 
and $high$ states. 
$^{\dagger\dagger}$ normalization parameters of $BMC$ 
component is in units of 
$L_{35}/d^2_{10}$ $erg/s/kpc^2$, where $L_{35}$ is the source luminosity in units of 10$^{35}$ erg s$^{-1}$, 
$d^2_{10}$ is the distance to the source in units of 10 kpc;  
$^{\dagger\dagger\dagger}$ for the {\it low state data}, we fit the spectrum using CASH statistic. In this case {\it the goodness-of-fit} is determined by Monte-Carlo simulations.
** Combined LS data accumulated during MJD 53023 -- 53134 \& 53310 -- 53316 (January -- May and November 2004 observations), 
indicated by *. }
\end{table*}

\subsubsection{Bulk Motion Comptonization model and its application to M101 ULX-1\label{bmc-results}}

The {\it Bulk Motion Comptonization} (BMC) model has  successfully fitted  the  M101 ULX-1 spectra
for all spectral states. Specifically, 
$Swift$/XRT spectra for band A ($red$) and band C ($blue$) fitted using the BMC 
model are presented 
in Figure~\ref{two_state_spectra} ($left$ panel). The plot highlights the 
significant spectral variability between these sets of the observations (see Figure~\ref{lc} for our definition 
of $Swift$/XRT count-rate bands, and Table 3 for the best-fit parameters). 
In Table 3 (at the bottom), we present the results of spectral fitting  $Swift$/XRT data of M101 ULX-1 using the
$phabs*bmc$ model. 
In particular, 
the LS$-$HS 
transition  
is related to 
 the photon index $\Gamma$ change from 1.4 
to 2.5 when the relatively low seed photon temperature $kT_s$ changes from 40 eV to 90 eV. 
Note the $bmc$ normalization varies by factor five, namely in the range of 
$2.9<N_{BMC}<15.6\times L_{35}/d^2_{10}$ erg s$^{-1}$ kpc$^{-2}$. 
While
the Comptonized (illumination) fraction is   quite low ($\log{A}<-4$ or $f\sim 10^{-4}$) 
{for all cases.}

As we have already pointed out  above,  Pence et al. (2001), Mukai et al. (2005), Kong et al. (2004) and Kong \& Di Stefano (2005)  analyzing the $Chandra$ data  investigated the spectral evolution of M101 ULX-1. We have also found a similar spectral behavior 
for the selected data set (see Table~\ref{tab:list_Chandra}) using our model. 
In particular, we have revealed that  M101 ULX-1 was in the {\it high state} during three outbursts: 
at 2000 (March and October); 
at 2004 July and  at 2004 December 30 -- 2005 January 1. 
The other $Chandra$ observations are related to the {\it low state} 
when the source is 
seen at the detection limit. The {low state} events of M101 ULX-1 covers  long time intervals: 
during 2004 January, March, May, November and December. Usually in the {\it low} state 
the X-ray luminosity of ULX-1 is about a factor 100 lower than that during the {\it high} state, when the peak bolometric 
luminosity (for assumed isotropic emission) is about $10^{41}$ ergs~s$^{-1}$. 

{In the $right$ panel of Figure~\ref{two_state_spectra} we demonstrate  
two representative $Chandra$ spectra for different states of M101 ULX-1. Data  taken for  2004 July 5 ($red$), 
which correspond to the  high state spectrum and for 2004 January -- May and November ($blue$)  which  correspond to   the  
low state spectrum. 
These spectra have been fitted  by a $phabs*bmc$ model  with the best fit parameters $kT_{s}=70$ eV  ($red$ solid line,  
for the {\it high} state) (HS) and $kT_{s}=45$ eV ($blue$ solid line, for the {\it low} state) (LS). 
We list the  best-fit spectral parameters 
in  
Table~\ref{tab:fit_table_Chandra}.
The shapes of these spectra related  to these two states,
are   different.
In the LS 
state  the seed photons  (with the lower $kT_{s}$ presumably related to lower mass accretion rate) are Comptonized more efficiently because the illumination fraction $f$ [or  $\log(A)$] is higher.
On the other hand in the HS state, these parameters, $kT_{s}$ and $\log(A)$  show an opposite behavior, namely 
 $\log(A)$ is lower for higher $kT_s$.  That means that 
a relatively small fraction of the seed photons, which temperature is higher because of the higher mass accretion rate in the HS than that in the LS, is  Comptonized.
}

We also evaluated  the blackbody radius $R_{BB}$ derived using a relation $L_{BB} = 4\pi R^2_{BB}\sigma T^4_{BB}$, where $L_{BB}$ is the luminosity 
of the blackbody and $\sigma$ is Stefan's constant. Assuming a distance D of 7.6 Mpc (as an upper estimate), the region associated with the blackbody has the radius $R_{BB}\leq 3\times 10^6$ km, which clear indicates  the IMBH 
presence in M101 ULX-1. In fact, $R_{BB}$ should be of order $10-30$ km for a Galactic BH of mass around 10 solar masses.  

It is worth noting that  our spectral model shows  very good performance throughout
all data sets. 
The reduced  $\chi^2_{red}=\chi^2/N_{dof}$ 
(where $N_{dof}$ is the number of degree of freedom) is  less or around 1.0 for the most of  the observations. For a small fraction (less than 3\%) of the spectra with high counting statistics
$\chi^2_{red}$ reaches 1.4. However, it never exceeds a rejection limit of 1.5.

\subsubsection{Evolution of X-ray spectral properties during spectral state transitions}

We have established common characteristics of the HS 
and LS spectral transitions of M101 ULX-1 (as seen in Fig. \ref{lc})  based on their
spectral parameter evolution of X-ray emission in the energy range from 0.3 to 7 keV using $Swift$/XRT 
and $Chandra$/ACIS data. 
In Figures~\ref{lc} 
we show the light curves highlighting 
the X-ray variability of the source. In Figure~\ref{chandra_lc}, {\it from top to bottom} we demonstrate  
an evolution of  
the seed photon temperature $kT_s$, 
the BMC 
normalization 
and  the spectral index $\alpha=\Gamma-1$ during 2004$-$2005 outburst transitions 
observed with $Chandra$/ACIS-S. 
The outburst 
phases  of the LS$-$HS 
transitions are marked by blue vertical strips. 

%
%

  \begin{figure*}
 \centering
    \includegraphics[width=17cm]{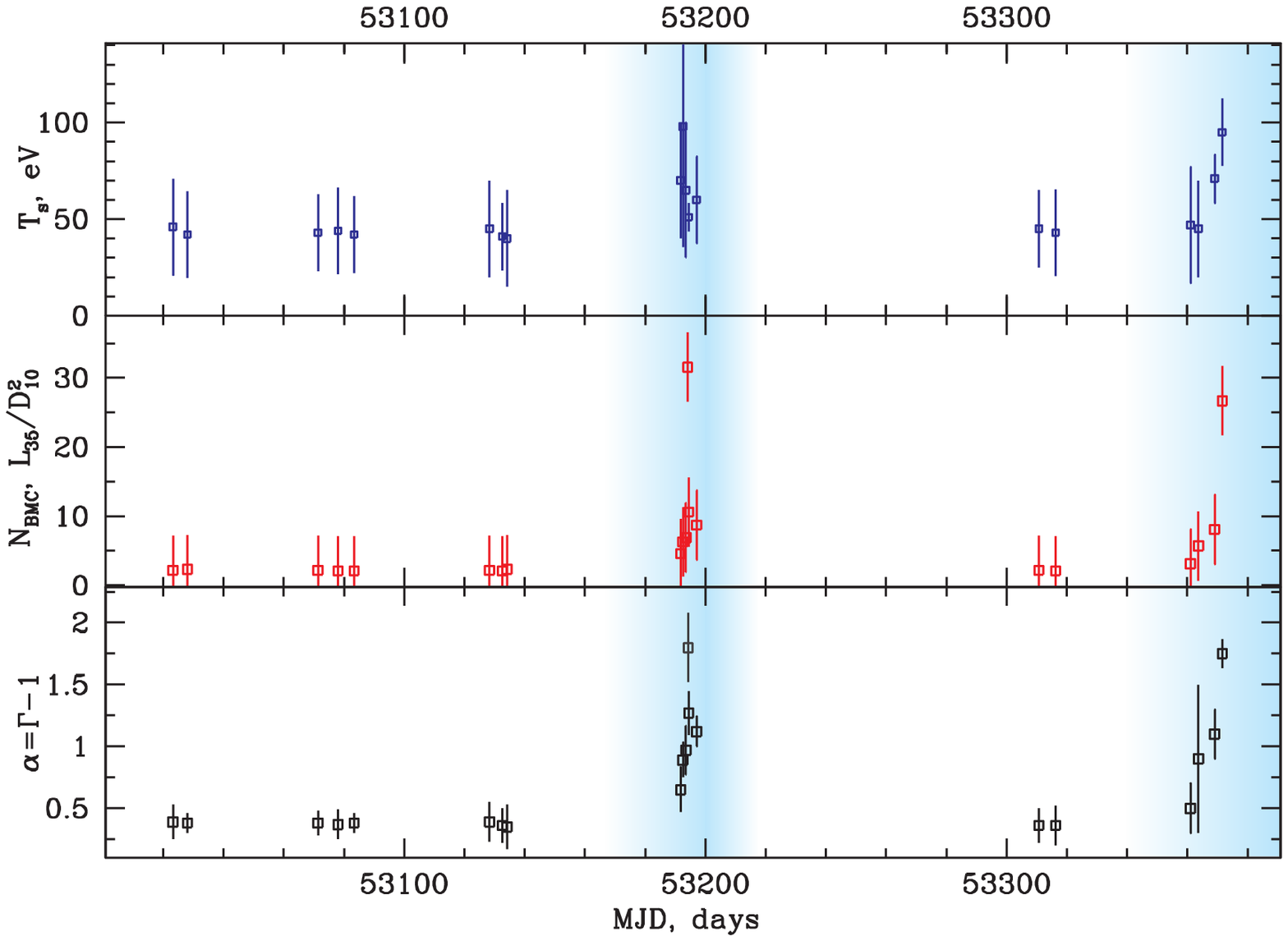}
      \caption{
{\it From Top to Bottom:}
Evolutions of the 
seed photon temperature $kT_s$ in eV, 
 the $BMC$ 
normalization 
and  the spectral index $\alpha=\Gamma-1$ during 2004$-$2005 
outburst transitions 
observed with $Chandra$/ACIS-S. 
The outburst 
phases  of the LS$-$HS 
transitions are marked by blue vertical strips. 
}
\label{chandra_lc}
\end{figure*}

During the rise phase and close to the peak of outburst, the softer emission 
[0.3$-$1 keV] 
dominates in the spectrum, which is associated with the seed photon temperatures $kT_{s}=40 - 60$ eV 
(see upper panel 
of Fig.~\ref{chandra_lc}). 
At the outburst  we detected  the maximum of the seed photon 
temperature $kT_{s}=100$ eV (see e.g. MJD=53194 point) along with the maximum 
of the normalization $N_{bmc}$. 
 Through the next 
days after outburst,  mass accretion rate $\dot M$ drops  by about a  factor of  ten (the 
BMC normalization parameter $N_{bmc}\propto\dot M$), $kT_{s}$
again drops to 60 eV when the source comes back  
its ``standard'' {\it low state}.  
In turn, a long ``standard'' {\it low state}  of M101 ULX-1 is associated with the low seed photon 
temperatures $kT_{s}=40$ eV (see e.g. MJD=53000 -- 53150 interval in $T_s$-panel of 
Fig.~\ref{chandra_lc}) and the low 
Comptonized fraction $f$ 
(see also Tables~\ref{tab:par_swift}$-$\ref{tab:fit_table_Chandra}).

From this plot we see that all spectral parameters correlate with each other during  the LS$-$HS transitions. 
In particular, the correlations of the photon index $\Gamma$ ($=\alpha+1$) versus BMC normalization $N_{BMC}$ 
are presented in Figure~\ref{saturation}, where 
{$blue$ 
triangles and 
$red$ 
squares 
are related  to  $Swift$ and $Chandra$   
data, 
respectively.
}
In Figure \ref{saturation}  we also show 
the photon index $\Gamma$ ($=\alpha+1$)
 monotonically increases from 1.3 to 2.8 with $N_{BMC}$ (proportional 
to $\dot M$) 
and saturates  at $\Gamma_{sat}=2.8\pm 0.1$ for  high values of $N_{BMC}$. 
One can see the strong saturation effect of the index $\Gamma$ versus  $N_{BMC}$. 


}
\section{Discussion \label{disc}}

Before  proceeding with the interpretation of the observations,
let us briefly summarize them as follows. (1) The spectral data of M101 ULX-1 are well fitted by the BMC model for all 
analyzed LS and HS spectra [see Figure~\ref{two_state_spectra} 
and  Tables~\ref{tab:par_swift}$-$\ref{tab:fit_table_Chandra}]. 
(2) The Green's function index of the BMC 
component $\alpha$ (or the photon index $\Gamma=\alpha+1$) rises and saturates with an increase of the BMC normalization (proportional to $\dot M$). The photon index saturation level 
of the BMC component is about 2.8 (see Figure~\ref{saturation}). 
}
}
\subsection{Saturation of the  index is a  signature of a BH  \label{constancy}}

Using our analysis of the evolution of $\Gamma$  in M101 ULX-1 
we have firmly established that {\it  $\Gamma$ saturates with the BMC-normalization $N_{BMC}$,  which is proportional
to $\dot M$}. ST09 give strong arguments that this $\Gamma$  saturation is 
a signature  of converging flow into  a BH. 

\begin{figure*}
 \includegraphics[width=12cm]{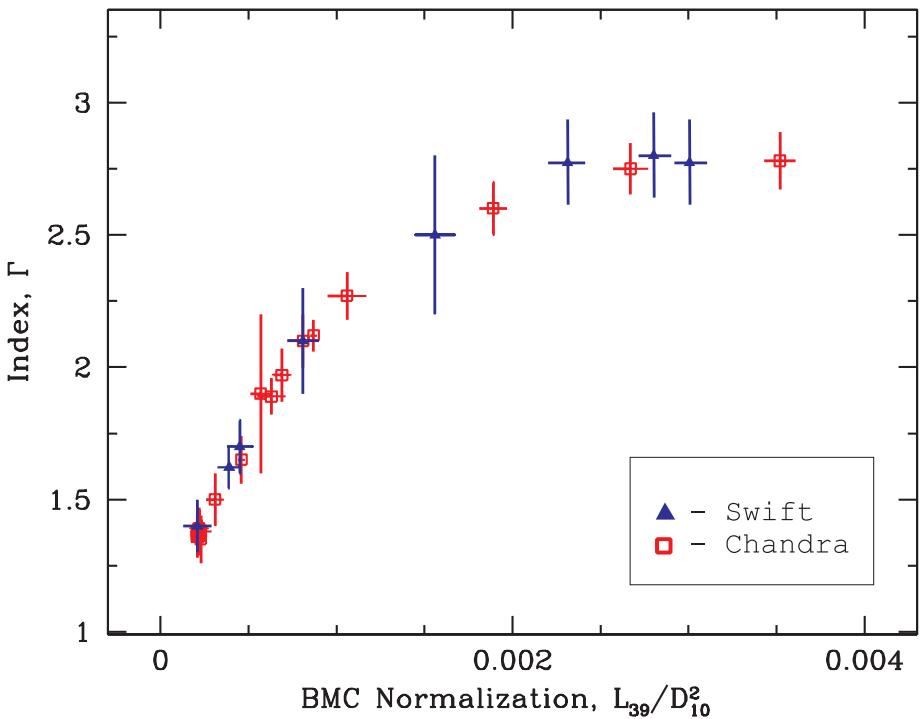}
   \caption{
Correlations of the photon index $\Gamma$ ($=\alpha+1$) 
vs  the BMC normalization $N_{BMC}$ (proportional to mass accretion rate) in units of $L_{39}/D^2_{10}$. 
$Blue$ triangles 
and $red$ squares 
are related  to  $Swift$ and $Chandra$ observations, respectively. 
}
\label{saturation}
\end{figure*}

\cite{tlm98} predicted that the transition layer (TL), the sub-Keplerian part of the accretion flow, should become more compact when 
$\dot M$ increases.  
For a BH case, \cite{tz98}, hereafter TZ98, obtain semi-analytically and later  \cite{LT99}, (2011), hereafter LT99 and LT11,
find, using Monte Carlo simulations,  that  $\Gamma$   saturates 
for high mass accretion rates.
Analyzing a number of Galactic BHs (GBHs)  ST09, \cite{tsei09}, \cite{ST10} and \cite{STS14} (STS14) confirm the  LT99-11 prediction   
that $\Gamma$ increases  and then it saturates with $\dot M$.
In  Figure~\ref{saturation}  one can see that  the  values of $\Gamma$ 
monotonically increase  from 1.3  and then they  finally saturate at a value of 2.8 for this particular source ULX-1 in M101.   

%
%

We observed the luminosity increase  along with 
the intrinsic softening of the spectrum lasting about three days.
When the luminosity drops,  we find the  spectral hardening as  a decrease of $\Gamma$ 
in agreement with the theoretical expectations  (see TZ98, LT99-11). This $\Gamma$  vs. $\dot M$ correlation found using  the M101 ULX-1 spectra are probably  driven by
 the same physical process that causes the  spectral 
evolutions seen in X-ray binaries due to  the  change of
$\dot M$. 
Moreover, we  argue that the X-ray observations of M101 ULX-1 reveal the {\it strong} index 
saturation vs $\dot M$ 
as  a signature of the converging flow 
(or BH presence) in this source (see ST09).
The index-{$N_{BMC}$ normalization (or $\dot M$) correlations found in a  number of GBHs
   allow us  to estimate a  BH mass  in M101 ULX-1 (see below \S 4.2).


%
%

\begin{table*}
 \caption{Parameterizations for reference and target sources}
 \label{tab:parametrization_scal}
 \centering 
 \begin{tabular}{lcccccc}
 \hline\hline                        
  Reference source  &       A       &     B     &   D  &    $x_{tr}$      & $\beta$  &  \\
      \hline
XTE~J1550-564 RISE 1998 & 2.84$\pm$0.08 &  1.8$\pm$0.3    &  1.0 & 0.132$\pm$0.004   &   0.61$\pm$0.02  \\
H~1743-322    RISE 2003 & 2.97$\pm$0.07 &  1.27$\pm$0.08  &  1.0 & 0.053$\pm$0.001   &   0.62$\pm$0.04  \\
4U~1630-472   & 2.88$\pm$0.06 &  1.29$\pm$0.07  &  1.0 & 0.045$\pm$0.002   &   0.64$\pm$0.03  \\
 \hline\hline                        
  Target source     &       A       &     B     &   D  &   $x_{tr} [\times 10^{-4}]$ & $\beta$ \\
      \hline
M101 ULX-1   & 2.88$\pm$0.06 &  1.29$\pm$0.07   & 1.0  &   4.2$\pm$0.2     &   0.61$\pm$0.03  \\
 \hline                                             
 \end{tabular}
 \end{table*}

\subsection{Estimate of BH mass in M101 ULX-1 \label{bh_mass}}


To scale the BH mass $M_{BH}$ of the target  source  (M101 ULX-1), we select   appropriate Galactic reference sources  
[XTE~J1550-564, H~1742-322 (see ST09) and 4U~1630-47 (STS14)] whose masses and distances  are known (see Table~\ref{tab:par_scal}), 
and also their BMC normalizations $N_{BMC}$.  
We can compare the index vs  $N_{BMC}$ (proportional to $\dot M$) correlations for these sources  with that of the target source   M101 ULX-1 (see Fig.~\ref{three_scal}). 
Note  that for all these sources the index saturation level  is  at the {almost} same value of $\Gamma$.   We  have used 
these three reference sources  for an additional cross-check of the BH mass evaluation {of M101 ULX-1}.

%
%
\begin{figure*}
 \includegraphics[width=12cm]{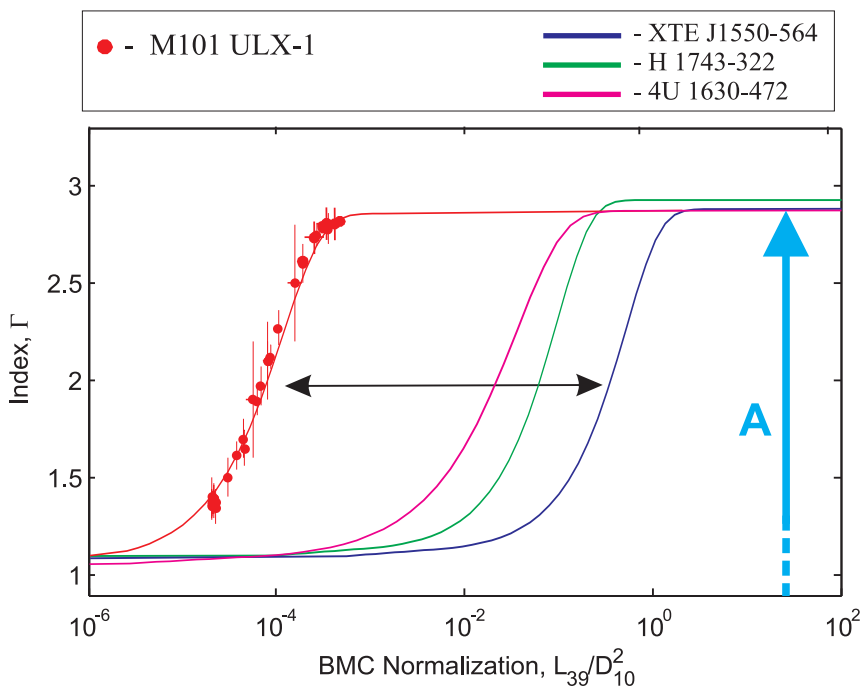} 
  \caption{
Scaling of the photon index $\Gamma$ versus the normalization $N_{BMC}$ for M101~ULX-1 (with $red$ 
points 
-- target source) and 4U~1630-472, 
XTE J1550-564 and H1743-322 (with $pink$, $blue$ and $green$ mark the reference sources), 
respectively. 
The {\it bright blue} vertical arrow 
schematically shows that the parameter A is the 
value of the index saturation level. 
The horizontal black arrow 
stresses the fact 
that the correlations of the  target and reference sources are similar.
The only difference is in terms of 
the BMC normalization, due to the different values of the $M_{BH}/D^2$ ratio.
}
\label{three_scal}
\end{figure*}

{
}




}

All correlation patterns   are  self-similar, showing the same index saturation level, which allows us to perform a reliable scaling.
The BH mass scaling technique is generally based on the parameterization of the $\Gamma-N_{BMC}$ correlation, that according to ST09 is fitted by a function


\begin{equation}
F(x)= A - (D\cdot B)\ln\{\exp[(1.0 - (x/x_{tr})^{\beta}))/D] + 1\}.
\label{scaling function}
\end{equation}
where $x=N_{BMC}$.


By fitting this function 
to the correlation pattern, we find a set of parameters A, B, D, $N_{tr}$, and $\beta$ 
that represent a best-fit form of the function $F(x)$ for a particular correlation curve. 
For $x\gg x_{tr}$, the correlation function F(x) converges to a 
constant value A. Thus, A is the value of the index saturation level, $\beta$ is the power-law index of the 
inclined part of the curve and $x_{tr}$ is a value at which index $\Gamma$ starts growing and $\beta$ provides 
the slope of the correlation. A parameter $D$ determines how smoothly the fitted function saturates to A. This function $F(x$) is widely used for a description of the correlation of $\Gamma$ vs $N_{BMC}$ 
[\cite{sp09}, ST09, \cite{ST10}, \cite{STS10}, STS14 and \cite{ggt14}]. 


The crucial assumption for this technique to be applied is that different reference sources show the same shape of the $\Gamma-N_{BMC}$ correlations and the only difference is in the ratio of
a BH mass to the squared distance, namely in the coefficient $M_{BH}/d^2$. 
 Figure ~\ref{three_scal} 
shows that a value of the parameter A 
(see {\it bright blue} vertical arrow) is 
almost the same for all scaling sources. In  other words, the best-fit parameter $A$
(within the limits of error bars)  is almost the same for all these sources. 
In particular, $A_{ULX}=2.8\pm 0.1$, $A_{1550}=2.84\pm 0.08$ 
and $A_{1743}=2.97\pm 0.07$ for M101 ULX-1, XTE J1550-564 and H 1743-322 respectively.
Furthermore, the black  horizontal arrow stresses that the correlations for  a pair of sources 
[e.g., M101 ULX-1 ($red$ line) and XTE~J1550-564 ($blue$ line)]  
are self-similar and the 
only difference is in the BMC normalization because of   
 the different values of the 
$M_{BH}/D^2$ ratio.
  
Thus, in order to obtain the BH mass of M101 ULX-1, one should shift along $N_{BMC}-$axis the related correlation of the reference source to the one of the target source 
(see Fig. \ref{three_scal}). This scaling technique provides a target BH mass value $M_t$:

\begin{equation}
M_t=M_r \frac{N_t}{N_r}
\left(\frac{d_t}{d_r}
\right)^2 f_G,
\label{scaling coefficient}
\end{equation}

\noindent where t denotes the target, r stands for the reference and the  geometric factor, by definition,  
$f_G=(\cos\theta)_r/(\cos\theta)_t$, the inclination angles $\theta_r$,  
$\theta_t$ and $d_r$, $d_t$ are distances to the reference and target sources respectively (see details in ST09). 
Note that the geometrical factor $f_G$ has to be considered when the accretion process is assumed to
occur in disk-like geometry, while it is close to 1 in case of  spherical accretion. 
Despite this uncertainty in the determination of $f_G$, we adopt the above formula for  $f_G$ in which $\theta\sim i$ 
if information on the system inclination angle $i$ is available (see Table \ref{tab:par_scal}).  

In Figure~\ref{three_scal} 
we plot the $\Gamma-N_{BMC}$ for M101 ULX-1 
points extracted using $Chandra$ and $Swift$ 
spectra 
 along with those for  the three reference patterns  [4U~1630-47 ($pink$), XTE J1550-564 ($blue$), H~1743-322 ($green$)]  
which  are similar  to the correlation found  for the target source.  
Scaling parameters for each of these pairs are presented in Table~\ref{tab:par_scal}. 

The target mass  for M101 ULX-1 can be estimated using the relation
\begin{equation}
M_t=C_0 {N_t} {d_t}^2 f_G 
\label{C0 coefficient}
\end{equation}
\noindent where 
$C_0=(1/d_r^2)(M_r/N_r)$ is the scaling coefficient for each scaling pair (target and reference sources), masses $M_t$ and $M_r$ are in solar units and $d_r$ is the distance to a particular reference source  measured in kpc.

We take values of $M_r$, $M_t$, $d_r$, $d_t$, and $\cos (i)$ from Table~\ref{tab:par_scal} 
and then  we obtain the lowest limit of the mass, using the best fit value of  $N_t= (4.2\pm 0.2)\times 10^{-4}$ taken at the begining of the index saturation  (see Fig. \ref{three_scal}) and measured
in units of $L_{39}/D^2_{10}$ erg s$^{-1}$ kpc$^{-2}$ [see Table \ref{tab:parametrization_scal}
 for values of the parameters of function $f(N_t )$ (Eq. 1)].
We estimate $C_0\sim 1.9, ~1.72~ 1,83$ 
for XTE J1550-564, H~1723-322 
{and 4U~1630-472}  respectively using $d_r$, $M_r$, $N_r$ presented by ST09. 
Then, using formula (\ref{C0 coefficient}), we obtain that $M_{ULX}\ge 3.4\times 10^4~M_{\odot}$ ($M_{ULX}=M_t$), 
assuming $d_{ULX}\sim 6.4$ Mpc~\citep{Shappee11} and  $f_G\sim1$ (inclinations for both objects are the same).  
To take account of the spread in the distance to M101, we have made the same estimates of $M_{ULX}$ assuming 
$d_{ULX}=7.4\pm 0.6$ Mpc~\citep{Kelson96} and derived  higher values $M_{ULX}$
$\ge 4.3\times 10^4~M_{\odot}$. 
All these results 
are summarized in  Table~\ref{tab:par_scal}. 

It is evident that the inclination of M101 ULX-1 system may be different from the inclination for the reference sources 
($i\sim 60-70^{\circ}$), therefore we take this  BH mass estimate for M101 ULX-1 as a lowest BH mass value  because 
that $M_{ULX}$ is reciprocal function of $\cos (i_{ULX})$
[see Eq.~\ref{C0 coefficient} taking into account that $f_G=(\cos\theta)_r/(\cos\theta)_t$ there]. 


The obtained  BH mass estimate is in agreement with a high bolometrical luminosity for M101 ULX-1 and $kT_s$ value which is in the range of 40$-$100 eV. 
In fact, a very soft spectrum is consistent with 
the relatively cold disk for ULXs that has also been considered as evidence for IMBHs (Miller et al. 2003, 2004;  Wang et al. 2004).


It is also important to note 
that Kong et al. (2004), based on the comparison between the observed temperature ($kT\le 100$ eV) and bolometric luminosity 
($L_{bol}\sim 10^{40-41}$ ergs s$^{-1}$) 
during the 2004 July outburst, 
obtained a similar estimate on BH mass of M101 ULX-1.  In fact, they obtained  that  BH  mass in M101 ULX-1, $M_{m101}$  is greater than  2800 $M_{\odot}$. Furthermore,  
Kong \& Di Stefano (2005) using the 90\% lower limits of the disk blackbody fits derived from the 2004 December outburst, 
estimated  $M_{m101}$ being  in the range of $1.3\times10^3 - 3\times 10^4 M_{\odot}$. 


%
%

\begin{table*}
 \caption{BH masses and distances.}
 \label{tab:par_scal}
 \centering 
 \begin{tabular}{llllcc}
 \hline\hline                        
      Source   & M$^a_{dyn}$ (M$_{\odot})$ & i$_{orb}^a$ (deg) & d$^b$ (kpc)  & M$_{scal}$ (M$_{\odot}$) \\
      \hline
XTE~J1550-564$^{1, 2, 3}$  &   9.5$\pm$1.1 &  72$\pm$5    &   $\sim$6           &   10.7$\pm$1.5$^c$ \\
H~1743-322$^4$     &   $\sim$11    &  $\sim$70    &   $\sim$10          &   13.3$\pm$3.2$^c$ \\
4U~1630--47$^5$    &       ...     &   $\leq$70   &   $\sim$10 -- 11    &   9.5$\pm$1.1     \\
M101~ULX-1$^{6, 7}$     & 3 -- 1000     &     ...      & (6.4$\pm$0.5)$\times 10^3$ & $\ge 3.2\times10^{4}$ \\
M101~ULX-1$^{7, 8}$     & 3 -- 1000     &     ...      & (7.4$\pm$0.6)$\times 10^3$ & $\ge 4.3\times10^{4}$ \\
 \hline                                             
 \end{tabular}
 \tablebib{  
(1) Orosz et al. 2002; 
(2) S$\grave a$nchez-Fern$\grave a$ndez et al. 1999; 
(3) Sobczak et al. 1999; 
(4) McClintock et al. 2007;  
(5) STS14;
(6) Shappee \& Stanek 2011;
(7) Mukai et al. 2005; 
(8) Kelson et al. 1996.
}
\tablefoot{ 
\\$^a$ Dynamically determined BH mass and system inclination angle, $^b$ Source distance found in literature, 
$^c$ Scaling value found by ST09. } 
 \end{table*}

\begin{figure}
   \resizebox{\hsize}{!}{\includegraphics{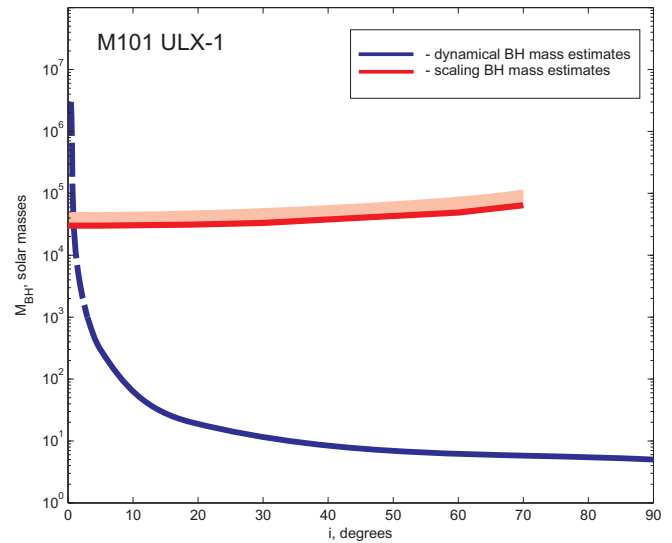}}
   \caption{
 Estimates of  BH mass $M_{BH}$ in M101 ULX-1. The $blue$ line represents 
 $M_{BH}$ versus the 
inclination angle $i$ estimated using the dymanical method for  HST data  [taken from \cite{Liu13}], while the 
$red$ line corresponds to our BH mass estimates based on X-ray 
data using the scaling technique (see Table \ref{tab:par_scal}).   
These two different methods  give  similar BH mass values  
with an assumption of small inclination angles.  
}
\label{dynam_scal_mass}
\end{figure}

\cite{Liu13}  report on optical observations  of M101 ULX-1 by Gemini/GMOS and they find that the system contains a Wolf-Rayet 
star with  an orbital period of 8.2 days.  The optical spectrum of the source is characterized by a broad helium emission 
line, including the He II 4686 \.{A} line.  Because of the absence of a broad hydrogen emission line the authors argue that 
the star must be a Wolf-Rayet (WR).  They propose the scenario that the intensities of the helium emission lines can be 
reproduced  by  the atmospheric model (see Hillier \& Miller  1998)  and the stellar mass is estimated to be 19 $ M_\odot$ 
based on the empirical mass-luminosity relation (Schaerer \& Maeder 1992 and Crowther 2007).  
Liu et al.  find the mass function  is about 0.18 $M_\odot$ for M101 ULX-1.  Suggesting different values of inclination angle $i$ they propose that this BH mass is likely 20 -- 30 $M_\odot$. 
In  
Figure \ref{dynam_scal_mass} 
we present   Liu's  BH estimate as  a function of inclination angle $i$.  The range of their BH mass estimates varies  from 5 to 1000 solar masses depending on inclination angle $i$.  For smaller $i$ a BH mass is higher (more than 1000 $M_\odot$) 
and  for $i\le 90^\circ$ it is about 5 solar masses. 

 Liu's evaluation of the  BH mass  (20-30 $M_\odot$) is too  low in comparison with our BH mass estimate  and also it is in contradiction with the  lower values of the soft seed photon temperature $kT_s$ (see discussion above).  In fact, for a BH of 20-30 $M_\odot$ the seed temperature $kT_s$ is expected to be around 0.5 keV (see ST09). \cite{Liu13} also point out that  these low temperatures of the seed (disk) photons $kT_s\sim 70$ eV combined with high luminosities ($> 10^{39}$ erg s$^{-1}$), which are observed in ULX-1 M101, complicate the interpretation of ULX-1 as a stellar-mass  BH.   

We  derived the bolometric luminosity from the normalization of the BMC model  between 
$7\times 10^{40}$ erg/s and $6\times 10^{41}$ erg/s  (assuming isotropic radiation). 
This  high luminosity  is difficult to achieve in a X-ray binary unless the accretor has a mass greater than  1000 $M_{\odot}$. 
While 
our luminosity estimate is higher than that for previous M101 ULX-1 
outbursts observed  
by XMM-$Newton$ in 2002 -- 2005 but  it is closer to 
that derived by Kong et al. (2004) (who used a combined power law plus $blackbody$ model). 
Note that on  average $L_x$ luminosity is lower than 
the bolometrical one  because the peak of the spectrum occurs at  relatively low photon energies ($E\sim 0.1$ keV). 

%
%

\section{Conclusions \label{summary}} 




 We have studied the low$-$high state transitions observed in  M101 ULX-1 using $Swift$ (2006 -- 2013) 
and $Chandra$ (2000, 2004 -- 2005) observations. We argued that the source spectra can be fitted by 
the BMC model for all observations. 
Our study reveals that the index$-$normalization (or $\dot M$) correlation observed in M101 ULX-1 is similar to those in GBHs. 
The photon index $\Gamma$ is  in the range $\Gamma = 1.3 - 2.8$. 
We have also estimated the peak bolometric luminosity, which is about $6\times 10^{41}$ erg s$^{-1}$.  

We applied the scaling technique based 
on the observed correlations to estimate $M_{BH}$  in M101 ULX-1. 
This technique is commonly and successfully 
applied to estimate BH masses of Galactic black holes.  In this work the scaling technique for  the first time  is applied  
to estimate  $M_{BH}$  in ULX. We obtain  values of 
{
$M_{BH}\sim (3.2-4.3)\times 10^4 M_{\odot}$,   
}  
which   are in a good agreement with that estimated by 
peak bolometric luminosity estimates.
{The low limit of this BH mass estimate is 
in agreement with optical results (see Liu et al., 2013) assuming the face-on system 
configuration in ULX-1
(see 
Fig.~\ref{dynam_scal_mass}). 
}  
Combining these estimates with  the inferred  low temperatures of the seed disk photons $T_s$ 
we can state  that 
the compact object of {ultra-luminous source} M101 ULX-1 is likely to 
{
be an intermediate-mass black hole with at least $M_{BH}> 3.2\times10^4 M_{\odot}$. 
}


\begin{acknowledgements}

This research was performed using  data supplied by the UK $Swift$ Science Data Centre at the University of Leicester. ES also thanks Phil Evans for useful scientific discussion. 
We appreciate editing the text  of the paper by Mike Nowak and Tod  Strohmayer. 
We also acknowledge the deep analysis of 
the  paper  by the referee and the editor.
\end{acknowledgements}

\newpage

\newpage


\end{document}